# PARTICLE ACCELERATION AND TRANSPORT AT THE SUN INFERRED FROM FERMI/LAT OBSERVATIONS OF >100 MEV GAMMA-RAYS

N. Gopalswamy[1], P. Mäkelä[2] and S. Yashiro[2]

ABSTRACT: The sustained gamma-ray emission (SGRE) events from the Sun are associated with an ultrafast (≥2000 km/s) halo coronal mass ejection (CME) and a type II radio burst in the decameter-hectometric (DH) wavelengths. The SGRE duration is linearly related to the type II burst duration indicating that >300 MeV protons required for SGREs are accelerated by the same shock that accelerates tens of keV electrons that produce type II bursts. When magnetically well connected, the associated solar energetic particle (SEP) event has a hard spectrum, indicating copious acceleration of high-energy protons. In one of the SGRE events observed on 2014 January 7 by Fermi/LAT, the SEP event detected by GOES has a very soft spectrum with not many particles beyond ~100 MeV. This contradicts the presence of the SGRE, implying the presence of significant number of >300 MeV protons. Furthermore, the durations of the type II burst and the SGRE agree with the known linear relationship between them (Gopalswamy et al. 2018, ApJ 868, L19). We show that the soft spectrum is due to poor magnetic connectivity of the shock nose to an Earth observer. Even though the location of the eruption (S15W11) is close to the disk center, the CME propagated non-radially making the CME flank crossing the ecliptic rather than the nose. High-energy particles are accelerated near the nose, so they do not reach GOES but they do precipitate to the vicinity of the eruption region to produce SGRE. This study provides further evidence that SGRE is caused by protons accelerated in shocks and propagating sunward to interact with the atmospheric ions.

## INTRODUCTION

Forrest et al. (1985) identified gamma-ray continuum from the Sun due to the decay of neutral pions produced by the interaction of energetic ions from the corona with the ions in the chromosphere. The primary characteristic of these sustained gamma-ray emission (SGRE) events is that the emission continues beyond the flare impulsive phase, sometimes up to almost a day. Neutral pions that result in SGRE require the precipitation ≥ 300 MeV protons. These particles are likely accelerated by shocks (Murphy et al. 1987) driven by coronal mass ejections (CMEs) as evidenced by the association of solar energetic particle (SEP) events and interplanetary type II radio bursts (Share et al. 2018; Gopalswamy et al. 2018). The CMEs are ultrafast (~2000 km/s) and halos, similar to CMEs that produce SEP events with ground level enhancement (GLE) implying the acceleration of particles to GeV energies (Gopalswamy et al. 2018). However, some SGRE events are associated with soft-spectrum SEP events indicating an apparent lack of >300 MeV protons (Winter et al. 2018). Gopalswamy et al. (2018) showed that the soft spectrum is a result of poor latitudinal connectivity of the shock nose to an Earth observer, so high-energy particles do not reach an Earth observer even though they reach the eruption site to produce SGRE. A similar argument was made by Gopalswamy et al. (2014) in explaining the lack of GLE events associated with ultrafast CMEs originating at higher latitudes. However, some GLE events do originate from higher latitudes (≥30º), but in these cases, the underlying CMEs seem to have been deflected toward the ecliptic, improving the connectivity, and hence enabling the detection of high-energy particles (Gopalswamy and Mäkelä, 2014).

The main reason behind the importance of the nose connectivity stems from the fact that the highest energy particles are accelerated over a small area near the shock nose; the area of particle acceleration increases with decreasing energy (Gopalswamy et al. 2021). The magnetic connectivity can be worsened by a CME deflection away from the ecliptic as in the case of the 2014 January 7 CME, which resulted in a soft-spectrum SEP event. Here, we examine various aspects of this CME that point to the importance of latitudinal connectivity.

## OBSERVATIONS AND RESULTS

The weak 2014 January 7 SGRE event was cataloged in Allafort (2018) and Ajello et al. (2021). The SGRE was associated with a fast full halo CME detected by the coronagraphs on board the Solar and Heliospheric Observatory (SOHO) and the Solar Terrestrial Relations Observatory (STEREO) missions. The event was associated with a large SEP event with >10 MeV proton

---

[1] Heliophysics Science Division, NASA Goddard Space Flight Center, Greenbelt, MD 20771, USA
[2] Department of Physics, The Catholic University of America, Washington DC 20064, USA

intensity of ~$10^3$ pfu detected by GOES and the particle detectors on board STEREO. An interplanetary type II radio burst was observed by the Radio and Plasma Wave instrument (WAVES) on board the Wind and STEREO: https://cdaw.gsfc.nasa.gov/CME_list/radio/waves_type2.html (Gopalswamy et al. 2019a). The type II burst also had a metric (m) component starting at 18:17 to 18:48 UT.

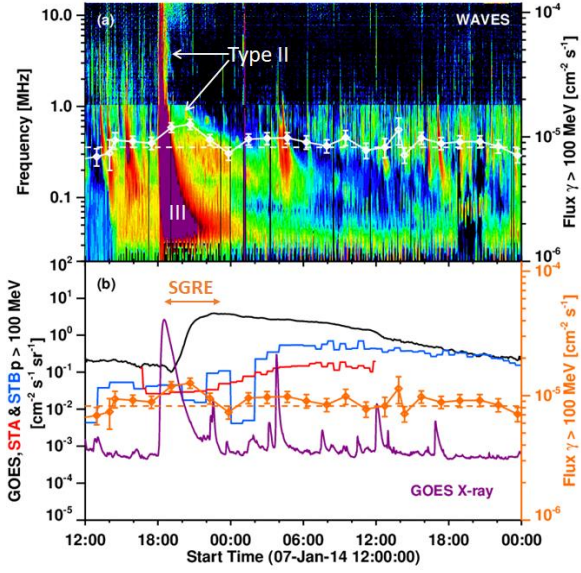

Fig. 1 (a) Wind/WAVES type II and type III radio bursts with Fermi/LAT >100 MeV light curve (white curve). (b) Proton intensity from GOES (black) and STEREO (STA, red; STB, blue) and the >100 MeV SGRE flux (orange).

Figure 1 shows the time evolution of the SGRE event using the light bucket method (Share et al. 2018) in comparison with that of the >100 MeV SEPs, IP type II burst, and GOES 1-8 Å light curve. The X1.2 flare starts, peaks, and ends at 18:04, 18:32, and 18:58 UT, respectively. The SGRE peak flux was ~$1.5 \times 10^{-5}$ cm$^{-2}$ s$^{-1}$ and the fluence was ~$4.77 \times 10^{-2}$ cm$^{-2}$. The SGRE duration ($T_{SGRE}$), measured from the peak time of the X1.2 flare to the mid time (23:03:45 UT) between the last Fermi/LAT signal data point and the one after that, is 4.53±0.77 hr. This duration is larger than that obtained using the maximum likelihood method (Allafort 2018; Ajello et al. 2021). To be consistent with our previous studies, we use the light bucket method. The IP type II burst starts around 18:33 UT and ends between 02:30 and 06:20 UT (mid time 04:25 UT), so the duration $T_{II}$ = 9.87±1.92 hr. As in other >3 hr SGRE events, the type II ending frequency is in the kilometric (km) domain: 200±90 kHz. The SGRE and type II burst durations agree with the relation $T_{SGRE}$ = (0.9±0.2) $T_{II}$ + (-0.8±1.9) reported in Gopalswamy et al. (2019b). The ending frequency ($f_e$) is also consistent with the relation $T_{SGRE}$ = (-0.07±0.01)$f_e$ + (25.5±2.7) indicating a strong shock far from the Sun. Such m-km type II bursts are characteristic of GLE events.

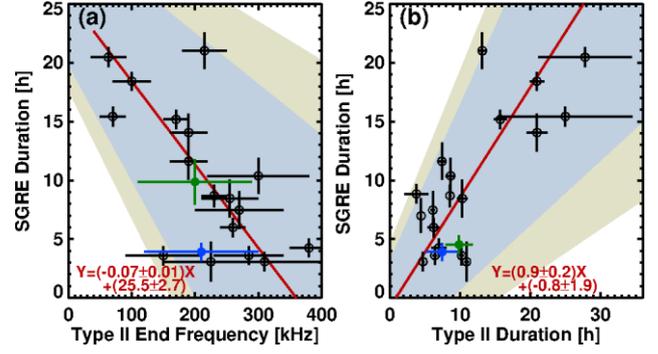

Fig. 2 (a) Scatter plot of SGRE duration with type II burst ending frequency (a) and duration (b) for 19 events with duration >3 h reported in Gopalswamy et al. (2019b). The green data point corresponds to the 2014 January 7 SGRE event. The blue data marks the 2014 September 1 SGRE event from a backside eruption (not included in the correlation). The shaded areas correspond to 95% and 99% confidence intervals. The 2014 January 7 event lies well within the 95% confidence interval. The red lines are linear fits to the data points (see text).

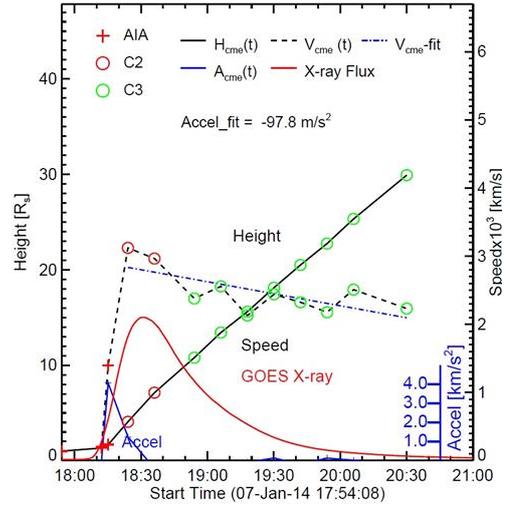

Fig. 3 Height ($H_{cme}$), speed ($V_{cme}$), acceleration ($A_{cme}$) as a function of time using GCS fit to SOHO, STEREO, and the Solar Dynamics Observatory's Atmospheric Imaging Assembly (AIA) images of the CME. The GOES flare light curve is shown for comparison.

**CME Kinematics**

The underlying CME is a fast halo CME originating from active region 11944 located at S15W11. The sky-plane speed of the CME was ~1830 km/s, which becomes 2246 km/s after a cone-model deprojection. The SDO, SOHO, and STEREO coronagraph data fit to the graduated cylindrical shell (GCS, Thernisien 2011) model gives a peak 3D speed of ~3100 km/s at 18:24 UT, and an average 3D speed of ~2400 km/s within the coronagraph field of view (Fig. 3). The SGRE fluence (F) - CME speed ($V_{cme}$) relation, log F = 6log($V_{cme}$/1000) – 2, gives F=1.91

cm$^{-2}$, which is larger than the observed value by a factor of 40, but within the typical scatter (Gopalswamy et al. 2019b). The initial acceleration peaked at ~4 km s$^{-2}$ (18:15 UT), which is consistent with the average acceleration (~1.86 km s$^{-2}$) obtained from the flare duration and average CME speed (Gopalswamy et al. 2016). Fig. 3 shows that the CME kinematics are similar to those of typical CMEs underlying GLE events. Note that the CME starts driving a shock indicated by the metric type II burst and attains its peak speed within the impulsive phase of the flare. One thing peculiar about the CME is that its nose is at position angle 231⁰, even though it originated close to the disk center (Fig. 4). The highly deflected motion has been reported before to be due to a combination of a coronal hole and the active region field not participating in the eruption (Gopalswamy et al. 2014; Möstl et al. 2015).

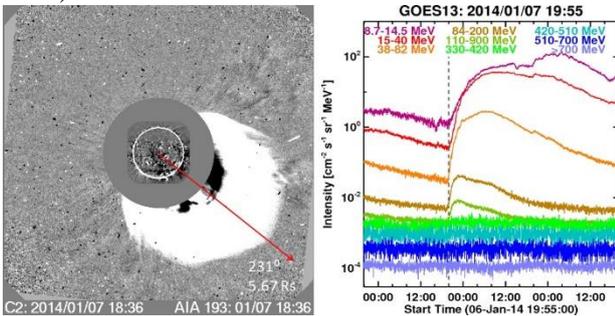

Fig.4. (left) LASCO image at 18:36 UT showing the nose at a height of 5.67 Rs and at position angle 231⁰. (right) GOES 13 proton intensity showing the SEP event starting at 19:55 UT above the background of a preceding event.

**The SEP spectrum**

Figure 4 shows the SEP intensity in various GOES proton channels and an integral channel at >700 MeV. The intensity rapidly falls off after the 38-82 MeV channel. In the wide channel (110-900 MeV) the event is clearly seen, but the contribution is mainly from the lower energies of the channel because there is barely discernible signal in the 330-420 MeV channel. Gopalswamy et al. (2016) reported the 10-100 MeV fluence power-law spectral index of the 2014 January 7 SEP event as 4.27 (see also Bruno et al. 2018), which is close to the average spectral index (4.89) of events associated with filament eruption (FE) CMEs, but much larger than that (2.68) of GLE events. It is clear that the soft spectrum results from the poor connectivity of the CME nose (Fig. 4) to an Earth observer. Thus, the 2014 January 7 SGRE provides evidence for the acceleration of >300 MeV protons by an ultrafast CME shock as in the case of other known SGRE events from higher latitudes (Gopalswamy et al. 2021).

**DISCUSSION AND SUMMARY**

The weak 2014 January 7 SGRE event is consistent with all the known relation of gamma-rays with CMEs, SEPs, and type II radio bursts. This event thus adds further evidence supporting the common shock origin of energetic particles producing SGRE events and those escaping into space detected as SEP events. The ultrafast halo CME and the long-enduring type II radio burst are characteristics of CMEs producing GLE in SEP events implying the required acceleration of high energy particles. Further investigation is needed to understand why the SGRE event is weak compared to other events with similar CME kinematics and SEP spectra but with a much stronger SGRE event (e.g., the 2011 March 7 event, see Gopalswamy et al. 2021).

**ACKNOWLEDGMENTS**

This work benefited from the open data policy of Fermi/LAT, SOHO, STEREO, SDO, GOES, and Wind teams. We thank H. Xie and N. Thakur for help with some figures. Work supported by NASA's LWS and GI programs.